\documentclass[a4paper, fleqn, svgnames]{cas-dc}
\usepackage[english]{babel}
\usepackage[numbers,compress]{natbib}
\usepackage[matha]{mathabx}
\usepackage{enumitem}
\usepackage{subfigure}      


\usepackage{capt-of} 
\usepackage{hyperref}
\hypersetup{
    colorlinks=true,
    citecolor=DarkRed,
    linkcolor=Navy,
    urlcolor=Navy,
    anchorcolor=black,
    linktocpage 
}
\usepackage{cleveref}
\graphicspath{{Figures/}}
\newcommand{\referencename}{ref.}
\newcommand{\refcite}[1]{\referencename\,\cite{#1}}

\newcommand{\Ns}{\text{N}_\sigma}
\newcommand{\Nt}{\text{N}_\tau}

\newcommand{\Nf}{\text{N}_\text{f}}
\newcommand{\clqcd}{CL\kern-.25em\textsuperscript{2}QCD}
\newcommand{\oqcd}{openQCD-FASTSUM}
\newcommand{\OCL}{OpenCL}
\newcommand{\bahamas}{\texttt{BaHaMAS}}
\newcommand{\crc}{CRC-TR\,211}
\crefname{figure}{Figure}{Figures}
\crefname{table}{Table}{Tables}
\crefname{equation}{Eq.}{Eqs.}
\crefname{section}{Section}{Sections}

\newcommand{\PartFunc}{\mathcal Z}
\newcommand{\Action}{\mathcal S}
\newcommand{\SGluon}{\Action_{\text{g}}}

\newcommand{\MuI}{\hat{\mu}_i}
\newcommand{\MuIpu}{\mu_i}
\newcommand{\MuIpuBar}{\bar{\mu}_i}
\newcommand{\RW}{Roberge-Weiss}

\newcommand{\OP}{Owe Philipsen}

\apptocmd{\MaketitleBox}{\vspace*{5mm}}{}{}

\begin{document}

\title[mode=title]{The chiral phase transition in the 3D Columbia plot~--~a scalable HPC workflow}
\tnotemark[1]
\tnotetext[1]{Preprint in its first version finalised on December 15, 2025}
\author[1]{Alessandro Sciarra}[auid=000,orcid=0000-0002-4608-1905]
\shorttitle{The chiral phase transition in the 3D Columbia plot}
\shortauthors{A.~Sciarra}
\ead{sciarra@itp.uni-frankfurt.de}
\credit{Conceptualization of the study, methodology, software development, writing}

\affiliation[1]{organization={ITP, Goethe Universität Frankfurt},
    addressline={Max-von-Laue-Str. 1},
    city={Frankfurt},
    postcode={60438},
    country={Germany}}



\begin{abstract}
    The nature of the chiral phase transition of QCD continues to represent a fundamental open problem in the study of strongly interacting matter.
    In recent years, significant progress has been achieved by exploiting systematic variations of theory parameters in regimes free of the sign problem.
    In this work, an open-source computational framework for large-scale lattice QCD investigations is presented, with particular emphasis on its application to extensions of previous studies at nonzero purely imaginary chemical potential.
    The framework is not developed specifically for this application: its underlying software components have been successfully employed for more than a decade in large-scale lattice QCD investigations.
    The focus is placed on the explicit documentation of a reproducible and scalable numerical workflow, covering all stages from data production on HPC systems to statistical analysis and final data interpretation.
    The presented framework integrates several publicly available software components, including GPU-accelerated simulation codes, automated job orchestration tools, and generic Monte Carlo analysis libraries.
    By bringing these previously operational components together and documenting their interfaces and interaction in a single framework, this work provides a ready-to-use computational blueprint for future large-scale studies of the QCD phase diagram.
    The corresponding software stack is fully released and designed to support systematic extensions of existing studies.
\end{abstract}

\maketitle


\section{Introduction}

Large-scale lattice QCD investigations of the QCD phase diagram represent a demanding computational challenge, requiring the coordinated execution of hundreds of long-running Monte Carlo simulations across heterogeneous HPC systems together with continuous monitoring, data synchronization, statistical analysis, and iterative parameter refinement.
In practice, modern studies in this field are no longer limited by the availability of simulation algorithms alone, but increasingly by the capability to integrate data production, workflow orchestration, and analysis into scalable and reproducible computational infrastructures.

The present work documents and systematizes an open-source workflow framework that has already been successfully employed over more than a decade in large-scale lattice QCD investigations, cf.\ e.g.\ \cite{Czaban:2015sas,Cuteri:2017gci,Cuteri:2020yke,Philipsen_2020,Cuteri_2021,Kaiser:2021rxe,DAmbrosio:2021awi}.
Rather than introducing a computational infrastructure developed specifically for the study discussed here, this article consolidates previously operational software components into a coherent and publicly available ecosystem.
The presented framework combines GPU-accelerated lattice QCD simulations, automated HPC job orchestration, intermediate data processing utilities, and recently-released generic Monte Carlo analysis tools into a unified computational pipeline.
Particular emphasis is placed on reproducibility, portability across heterogeneous HPC facilities, interface standardization between workflow layers, and extensibility towards future large-scale studies.
While the individual components are publicly released and documented separately, their role as parts of a single end-to-end computational ecosystem has so far not been presented in a unified form.
The present work therefore provides a common reference point for the complete workflow, making explicit how simulation, resource management, data handling, and statistical analysis interact during a large-scale lattice QCD campaign.

The underlying physics motivation concerns the phase structure of Quantum Chromodynamics (QCD) as a function of temperature and baryon chemical potential, which remains one of the central open problems in the theory of strong interactions.
Despite sustained theoretical and numerical efforts over several decades (see e.g.~\cite{AARTS2023104070} for a recent review), a comprehensive and quantitatively controlled understanding of the QCD phase diagram is still lacking.
This difficulty is rooted in the intrinsically non-perturbative nature of the strong interaction at hadronic scales, which renders analytic approaches ineffective in the regimes of primary phenomenological interest and makes first-principles approaches like lattice simulations indispensable.

Lattice QCD has provided firm results for the thermal properties of strongly interacting matter at vanishing baryon chemical potential, where Monte Carlo simulations are not hindered by the sign problem.
However, extending these results to nonzero baryon density remains notoriously challenging.
As a consequence, significant effort has been devoted to studying the dependence of the thermal transition on QCD parameters that can be varied without encountering a sign problem, such as the quark masses, the number of quark flavours, the lattice spacing, and purely imaginary chemical potential.
These investigations have proven to be a powerful tool for constraining the global structure of the phase diagram.
For an overview of recent progress, refer for instance to~\cite{Cuteri:2023evl} and the references therein.

In particular, the order of the chiral phase transition in the massless limit has long been recognized as a key theoretical question, with far-reaching implications for the QCD phase diagram at physical quark masses and for the possible existence and location of critical points at nonzero density.
Over the years, a variety of strategies have been developed to address this problem, combining numerical simulations with scaling analyses and universality arguments.
Among these, the approach introduced in \refcite{Cuteri_2021} provided a novel and systematic framework to investigate the nature of the chiral transition by exploiting controlled variations of theory parameters in regimes free of the sign problem.
A possible extension to purely imaginary chemical potential of such investigation is presented in this work.

The structure of this paper reflects both the physics motivation and the computational perspective.
\cref{sec:Columbia} presents the idea of extending a previous investigation of the Columbia plot to purely imaginary chemical potential and discusses the corresponding physical scenarios motivating the numerical study.
The core contribution of the manuscript is then presented in \cref{sec:computational}, where the previously operational but now systematically documented scalable computational workflow and its software ecosystem are described in detail, including simulation management, data handling, statistical analysis, software engineering practices, and reproducibility aspects.


\section{The 3D Columbia plot}\label{sec:Columbia}

\begin{figure*}[t]
    \vspace{1mm}
    \def\dx{0.48}
    \centering
    \subfigure[]{\label{fig:3D-1st}\includegraphics[width=\dx\textwidth]{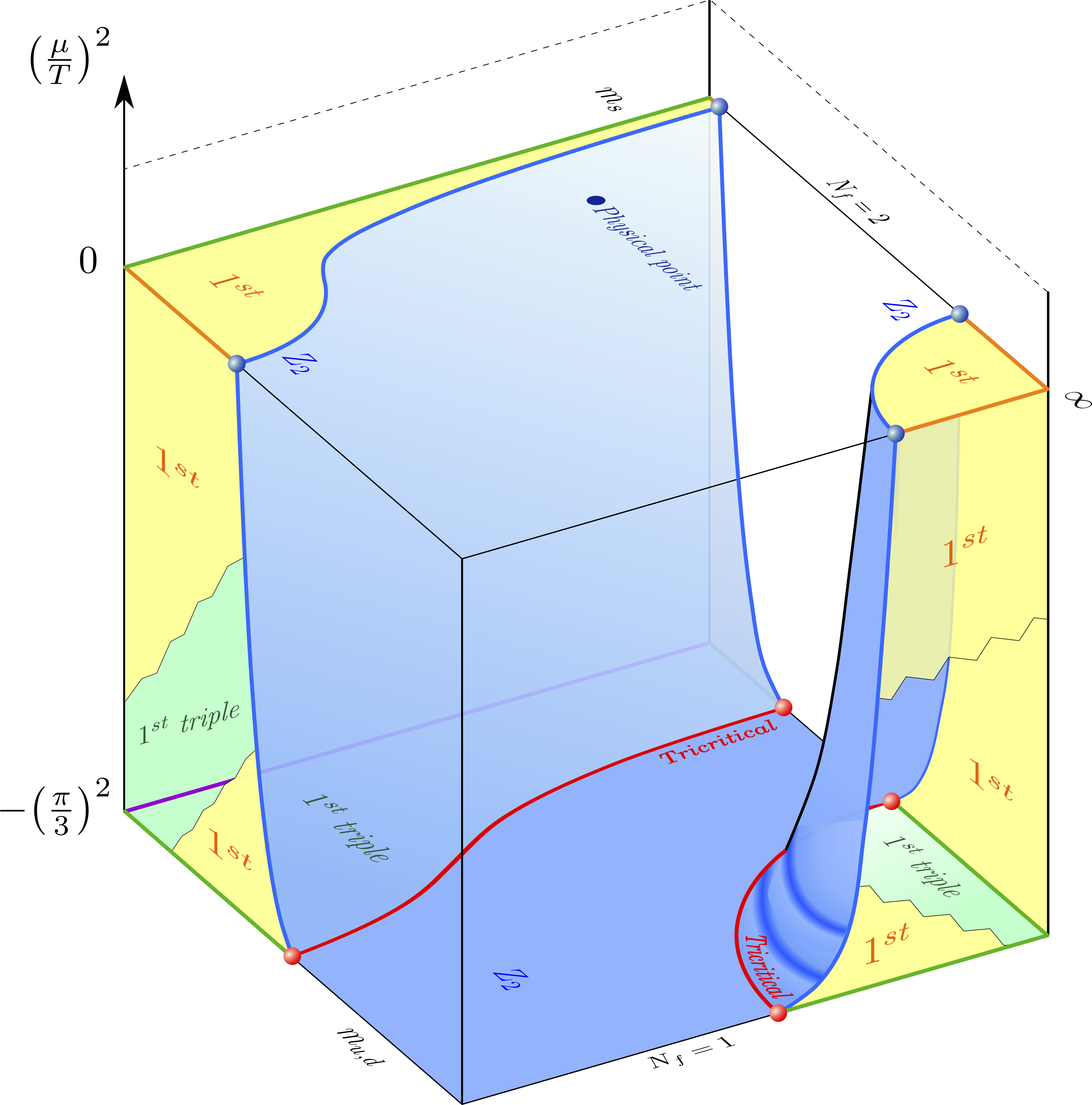}}\hfill
    \subfigure[]{\label{fig:CP}\includegraphics[width=\dx\textwidth]{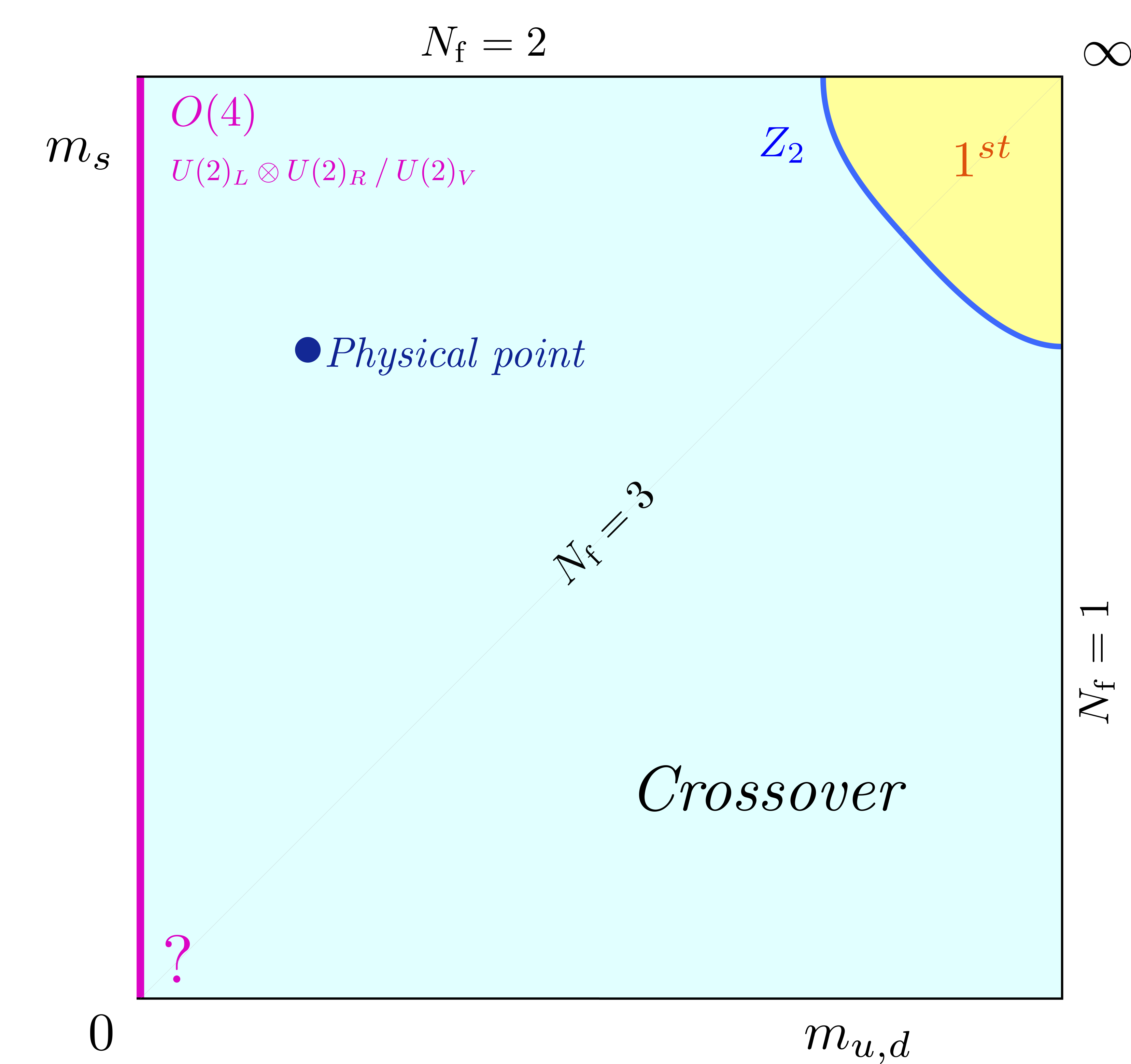}}
    \caption{
        Known sketches of the Columbia plot and its 3D extension.
        Sketch \protect\subref{fig:3D-1st} was taken from \refcite{Philipsen_2020} and depicts the scenario realized on coarse lattices.
        Sketch \protect\subref{fig:CP} was taken from \refcite{Cuteri_2021}; there it has been shown that the chiral first-order region at zero baryon chemical potential is a lattice artefact for $\Nf=2$ and $\Nf=3$ which vanishes in the continuum limit.
    }
    \label{fig:known-CP}
\end{figure*}

The implications of introducing a purely imaginary chemical potential $\mu=i\MuIpu$ with $\MuIpu\in\mathbb{R}$ in QCD is well understood~\cite{Roberge:1986mm}.
In particular, because of the absence of any sign problem, QCD can be simulated on the lattice at any value of \mbox{$\MuIpu\neq0$} and this technique has been extensively used in the last decades (see~\cite{sym13112079} for a recent review and references therein).

Using the same lattice setup used in \refcite{Cuteri_2021}, it is possible to include a nonzero purely imaginary chemical potential $\MuI=a\MuIpu$ in the temporal gauge links,
\begin{equation}
    \tilde{U}_{\lambda,\nu}=\left\{
    \begin{aligned}
        U_{\lambda,\nu} &\qquad\nu\in \{1,2,3\rbrace \\
        e^{i\MuI}U_{\lambda,\nu} &\qquad\nu=4
    \end{aligned}
    \right.\quad,
\end{equation}
where $\lambda$ and $\nu$ refer to lattice sites.
The partition function using the standard Wilson gauge action $\SGluon$ and the unimproved staggered discretization of dynamical fermions then reads
\begin{equation}
    \PartFunc(T,\MuI)=\int \mathcal{D}U \;\bigl(\det D[U,\MuI]\bigr)^{\Nf/4} \,e^{-\SGluon[U]}\;.
\end{equation}
$\SGluon$ includes the lattice gauge coupling $\beta=6/g^2$, on which the lattice spacing $a$ depends.
Together with the Euclidean time extent $\Nt$ of the lattice they specify the temperature as
\begin{equation}
    T=\frac{1}{a(\beta)\:\Nt}\;.
\end{equation}

In \refcite{Cuteri_2021} it has been found that the chiral first-order region that exists for $\Nf=2$ and $\Nf=3$ in the Columbia plot for coarse lattices is merely a lattice artefact which disappears in the continuum limit, where the chiral transition is of second order as depicted in \cref{fig:CP}.
Introducing a third (vertical) axis in the Columbia plot for the square of the chemical potential, it is possible to sketch the so-called 3D Columbia plot.
Below the $\mu=0$ plane, due to the periodicity of the partition function, it is enough to consider the space portion defined by $-(\pi/3)^2 \le (\mu/T)^2 \le 0$, where $(\mu/T)^2=-(\pi/3)^2$ is often referred as the \RW\ plane.
On coarse lattices, $Z_2$ surfaces connect the $Z_2$ lines in the Columbia plot to the tricritical lines in the \RW\ plane.
As the chiral first-order region in the Columbia plot shrinks to then disappear approaching the continuum limit, it is natural to wonder what happens at nonzero purely imaginary chemical potential.

\begin{figure*}[t]
    \vspace{5mm}
    \def\dx{0.48}
    \centering
    \subfigure[]{\label{fig:3D-seculation-1}\includegraphics[width=\dx\textwidth]{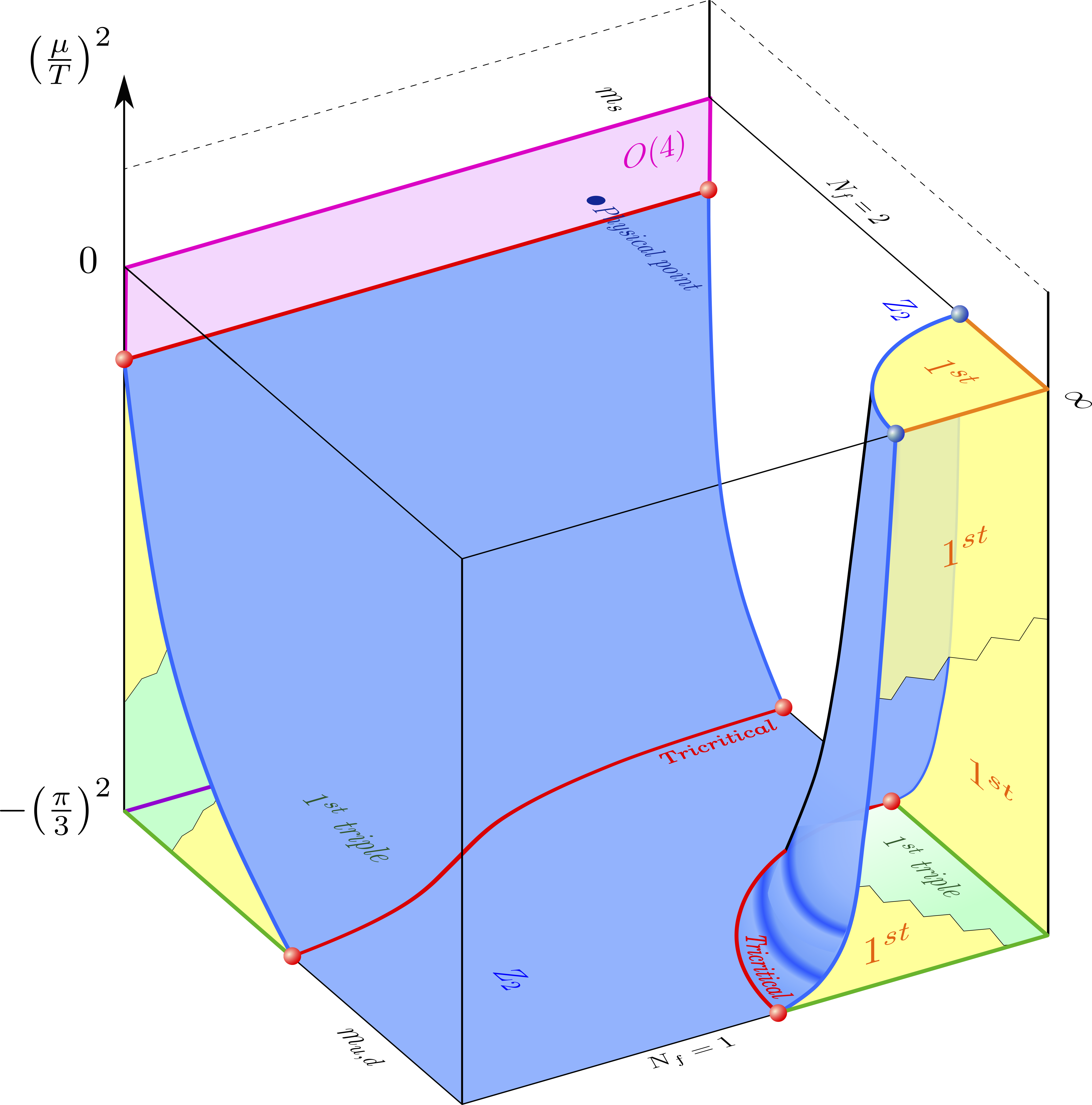}}\hfill
    \subfigure[]{\label{fig:3D-seculation-2}\includegraphics[width=\dx\textwidth]{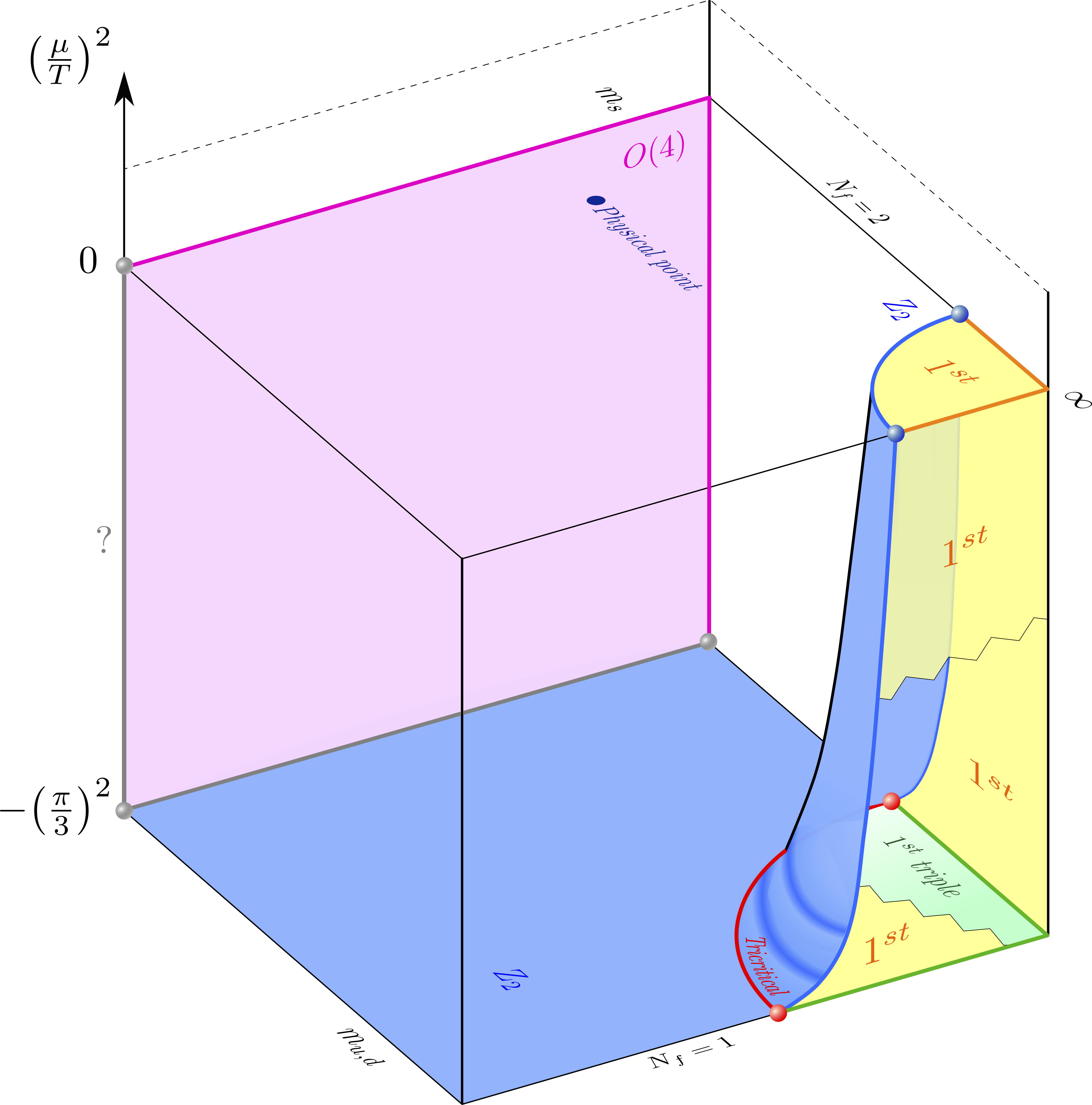}}
    \caption{
        Possible realizations of the 3D Columbia plot in the continuum limit~\cite{sciarra_2024_10599969}.
        Sketch \protect\subref{fig:3D-seculation-1} had already been suggested in \refcite{sciarra_PhD} as a speculation, while \protect\subref{fig:3D-seculation-2} represents the most natural scenario, given how the chiral first-order region seems to shrink approaching the continuum limit.
        Grey lines and points denote a possibly more complex type of phase transition, still to be determined.
        For instance, it is still an open question how the \RW\ transition across different sectors interplays with the thermal transition along the $m_{u,d}=0$ line in the \RW\ plane.
    }
    \label{fig:new-3D}
\end{figure*}

The easiest possibility to clarify this aspect, which emerged during the final stages of \refcite{Cuteri_2021}, consists in repeating that study at an intermediate value of purely imaginary chemical potential $\MuIpu$, such that \mbox{$-(\pi/3)^2 < (i\MuIpu/T)^2 < 0$}.
It is worth noting that setting $\mu$ to the \RW\ critical value would make the numerical investigation significantly more challenging, because of the complex phase structure of the theory due to the additional \RW\ symmetry.
In fact, in the \RW\ plane, varying the value of $\beta$ at a fixed value of $am$, $\Nf$ and $\Nt$ would mean to move along the boundary between two \RW\ sectors and this transition is known to be of first order for $\beta>\beta_c$ and a crossover for $\beta<\beta_c$~\cite{PhysRevD.67.014505,DEFORCRAND2002290}, where $\beta_c$ defines the location of the so-called \RW\ end-point.
How the presence of such additional phase transition would influence the signal of the kurtosis of the chiral condensate as function of $am$ on finite lattices should be numerically explored and it is difficult to predict.

Given the previous findings at $\mu=0$, also the $Z_2$ chiral surface in \cref{fig:3D-1st} will move towards the $m_{u,d}=0$ plane in the continuum limit at small values of $\MuIpu$.
If a first-order region at purely imaginary chemical potential survives in the chiral limit, this would be a region of first-order triple points, due to the coexistence of three phases at the critical temperature at $m_{u,d}=0$.
Such a region would then need to terminate in a tricritical line, that would be entirely at \mbox{$\MuIpu\neq0$}, because of the outcome of \refcite{Cuteri_2021}.
A possibility is reported in \cref{fig:3D-seculation-1}, which would still admit a first-order triple nature of the \RW\ end-point.
However, it might also be possible (or even natural) to expect that the entire first-order chiral region in the 3D Columbia plot disappears in the continuum limit and \cref{fig:3D-seculation-2} scenario is realised.
In this case, some aspects would require further investigation to be clarified.
For instance, the type of the phase transition in the chiral limit for $\Nf=3$ at any value of the imaginary chemical potential between 0 and the \RW\ value or for $\Nf=2+1$ in the \RW\ plane would remain to be determined.
This is depicted by grey lines and dots in \cref{fig:3D-seculation-2}.

To distinguish between \cref{fig:3D-seculation-1} and \cref{fig:3D-seculation-2} is not expected to be trivial.
Choosing the \RW\ critical value of the imaginary chemical potential would probably be the most sound approach, but it might be numerically too demanding, as already mentioned.
The best viable alternative for a first exploratory study is therefore to fix a value $\MuIpu=\MuIpuBar$ close enough to the \RW\ value, such that it may be reasonably inferred what could happen to the first-order triple chiral region in the \RW\ plane in the continuum limit.

Even such a first exploratory study would imply a numerical cost comparable to that of \refcite{Cuteri_2021}, which required simulations at more than 600 different $\beta$-values across the phase space.
Although the previous investigation enables a considerably more educated choice of simulation parameters, the study would still involve a large number of strongly interconnected simulations together with continuous monitoring, synchronization, intermediate analysis, and iterative refinement of the parameter scan.
Managing such a workflow manually rapidly becomes impractical at this scale and strongly motivates the adoption of dedicated computational infrastructures of the type discussed in the next section.


\section{Computational strategy and software framework}\label{sec:computational}

Large-scale finite-temperature lattice QCD studies typically require coordinated workflows spanning simulation, resource management, data synchronization, statistical analysis, and iterative parameter refinement across heterogeneous HPC environments.
The proposed extension of the study in \refcite{Cuteri_2021} to nonzero purely imaginary chemical potential provides a representative example of such a computational campaign.
Importantly, the workflow described in this section is not specific to this proposed extension.
Its underlying components and operational principles have been developed and successfully used over more than a decade in large-scale lattice QCD projects, including the investigation of \refcite{Cuteri_2021}.
The purpose of this section is therefore to document the resulting ecosystem as a whole, rather than to introduce a workflow developed for the particular physics case considered in \cref{sec:Columbia}.

From a conceptual perspective, the decomposition of the workflow into simulation, orchestration, monitoring, and analysis layers follows a standard paradigm commonly encountered in computational science.
However, the concrete realization of such a framework for large-scale lattice QCD studies poses highly domain-specific challenges related to long Monte Carlo trajectories, correlated statistical data, adaptive parameter scans, heterogeneous software stacks, and HPC walltime limitations.

The workflow documented in this work addresses these aspects through the integration of modular open-source software components with clearly separated responsibilities and standardized interfaces.
This design enables scalable execution of large simulation campaigns in a reproducible and largely automated manner, significantly reducing the operational complexity of investigations that would otherwise be impractical to manage manually.

\begin{table}[t]
    \centering
    \begin{tabular}{l@{\hspace{0mm}}r}
        \toprule
        \textbf{Aspect} & \textbf{Quantitative evidence} \\
        \midrule
        Period of operational use
        & Since 2014\\
        Large-scale lattice QCD studies
        & Over 10\\
        Monte Carlo simulations
        & $\mathcal{O}(10^3)$ per study\\
        Distinct $\beta$-values typically explored
        & $\mathcal{O}(10^2)$ per study\\
        Typical simulation duration
        & Some weeks\\
        Maximum simulation duration
        & $\mathcal{O}(1\;\mbox{year})$\\
        HPC systems / facilities
        & 5\\
        LQCD simulation codes integrated
        & 2\\
        \bottomrule
    \end{tabular}
    \caption{%
        Quantitative characteristics illustrating the scale, operational history, and modularity of the presented computational workflow.
        The values summarize previous large-scale lattice QCD investigations in which the underlying software ecosystem has been employed and characterize the practical scope of the workflow.
    }
    \label{tab:workflow-scale}
\end{table}

\begin{figure*}
    \centering
    \includegraphics[width=\textwidth]{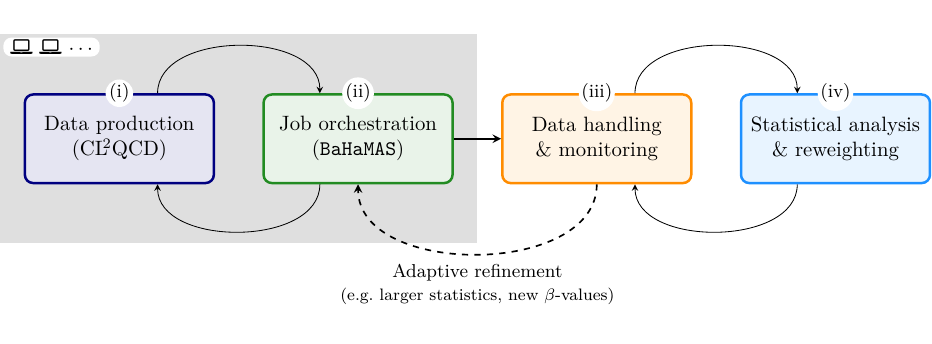}
    \caption{
        Schematic representation of the computational workflow illustrating its iterative and interconnected structure.
        The four stages introduced in the text are grouped into two tightly coupled cycles: data production and job orchestration \ref{item:raw}--\ref{item:job}, and data handling together with statistical analysis \ref{item:sync}--\ref{item:analysis}.
        Within each pair, bidirectional arrows indicate their continuous interplay during large-scale simulations, where job management and monitoring are intrinsically linked to data generation, and data processing is repeatedly interleaved with statistical analysis.
        The dashed arrow highlights the global feedback mechanism, through which intermediate analysis results inform subsequent simulation parameters, enabling adaptive refinement of the parameter space.
        The shaded region emphasizes that the data production and orchestration stages may be distributed across multiple HPC systems and simulation codes, whose outputs are subsequently synchronized and processed within a unified analysis environment.
    }
    \label{fig:workflow}
\end{figure*}

The overall workflow can be decomposed into four logically distinct layers: \ref{item:raw} data production, \ref{item:job} job orchestration and resource management, \ref{item:sync} data processing and monitoring, and \ref{item:analysis} statistical analysis and post-processing.
In practice, these layers are tightly interconnected and form a continuous pipeline rather than independent stages.
In particular, data synchronisation and preliminary data analysis are often performed many times while raw data is being produced until the desired statistics is collected.
This serves both the purpose to start new simulations (e.g.\ on larger lattice volumes or at new $\beta$-values) as soon as possible and to keep under control scanning resolution (e.g.\ in $\beta$).
For clarity, the workflow is schematically summarized in \cref{fig:workflow}, where the interplay between the different stages and their iterative nature are explicitly illustrated.
The individual steps are described below in the logical order in which they are typically executed in a large-scale lattice QCD project.
The scale of the computational campaigns for which this workflow has been employed is substantial.
All previous investigations~\cite{Czaban:2015sas,Cuteri:2017gci,Cuteri:2020yke,Philipsen_2020,Cuteri_2021,Kaiser:2021rxe,DAmbrosio:2021awi} made use of the underlying software ecosystem and involved large ensembles of Monte Carlo simulations, typically running for several weeks to many months on different supercomputing systems.
Some quantitative characteristics of these campaigns are summarized in~\cref{tab:workflow-scale}, which provides a direct indication of the scale at which the workflow has been exercised in production and complements the component-level description given below.
The workflow is therefore designed around the practical requirements of numerical campaigns in which hundreds of simulations must be coordinated and repeatedly revisited as intermediate analysis becomes available.

\begin{enumerate}[label={(\roman*)}, leftmargin=\parindent]
    \item\label{item:raw} \textbf{Data production.}
        For raw data production the publicly available, last release \texttt{v1.1} of the \OCL-based code \clqcd~\cite{cl2qcd} can be used out of the box.

        This software implements the Rational Hybrid Monte Carlo (RHMC) algorithm for unimproved rooted staggered fermions and supports simulations at purely imaginary chemical potential through a modification of temporal gauge links.
        Its design targets both CPU and GPU architectures, taking advantage of the OpenCL paradigm.
        Although the software is highly optimised for AMD GPUs, it also achieves competitive performance on NVIDIA hardware, enabling efficient large-scale Monte Carlo sampling across heterogeneous systems.

        From an algorithmic perspective, the RHMC implementation provides sampling of the fermion determinant via rational approximations, with tunable parameters controlling molecular dynamics integration as well as Dirac operator inversion accuracy and stability.
        \clqcd\ offers usage of multiple pseudofermions~\cite{Kennedy:2007}, which turns out to be advantageous also not so close to the chiral limit~\cite{Cuteri_2021}.
        These parameters can be systematically reused from the $\mu=0$ study, allowing for an informed initial parameter tuning and avoiding a complete re-optimization.

    \item\label{item:job} \textbf{Job orchestration and HPC integration.}
        The proposed study requires the submission, monitoring, and resumption of hundreds of simulations on one or more supercomputing systems.
        This can be entirely taken over by \bahamas\ in its new version \texttt{v0.5}~\cite{BaHaMAS}.

        This tool provides an abstraction layer for batch system interaction, enabling automated submission, monitoring, and resumption of large ensembles of independent Markov chains.
        The relevance of such a tool becomes evident when considering that the simulation of each set of phase-space parameters requires between few weeks to many months of simulation time on a typical supercomputer, which in turn typically impose maximum walltime limits of the order of a few days at most.
        Efficient monitoring and automated resumption of simulations are therefore essential in such large-scale projects (cf.\ e.g.\ \bahamas\ database report functionality~\cite{BaHaMAS-proc}, which can be regularly run in background).

        Moreover, the computational overhead introduced by the orchestration layer is negligible compared with the duration of the simulations it manages: job submission, status monitoring, and resumption operate at the level of individual batch-system operations, while the associated Monte Carlo campaigns run for weeks to months.

        \bahamas\ further supports parameter scans over the phase space including the imaginary chemical potential and allows optional containerized execution, improving portability across heterogeneous HPC environments.
        Parameters to be varied in different simulations ($\Nf$, $\Nt$, $\Ns$, $m_q$, $\MuIpu$) are extracted directly from the folder structure, ensuring a single source of truth for input parameters and reducing the risk of inconsistencies.

        From a workflow perspective, \bahamas\ acts as the central coordination unit, ensuring fault tolerance and efficient utilization of computational resources.
        In particular, this layer decouples physics input from system-specific execution details, which is essential when deploying the same study across multiple HPC facilities.
        As an example of the power of \bahamas\ as interface, it is worth remarking that also the highly parallel CPU software \oqcd~\cite{openQCD-FASTSUM} can be handled by \bahamas.
        Raw data produced in the software-specific output format (e.g.\ either by \clqcd\ or \oqcd) are always stored in a unified format by \bahamas, which enables the analysis software to accept raw data independently from which codebase has been used to produce them.
        This design effectively enforces a separation of concerns between data production and data analysis layers, while introducing a standardized interface for data exchange.
        As a consequence, simulation codes can be interchanged without affecting downstream analysis, provided that they comply with the common data format.
        This abstraction significantly improves the modularity of the workflow and facilitates both reproducibility and cross-validation across different codebases.

    \item\label{item:sync} \textbf{Data handling and intermediate processing.}
        While handling of simulations on the supercomputer is taken over by \bahamas, operations like synchronization of raw data and checkpoints, extracting information on the fly from running simulations, preliminarily fitting elaborated data, plotting histograms or interpolating data to properly start new simulations can be performed with a suite of utility scripts, which is currently released as a preliminary version \texttt{v0.1}~\cite{ScriptUtilities}.

        This intermediate layer is essential for adaptive refinement of the parameter space, reducing computational overhead.
        In practice, this stage together with the analysis software provides the feedback loop of the entire workflow, enabling decisions on where to increase statistics or refine parameter scans based on partially accumulated data.

        Although simple tasks like synchronising data between different servers, run the analysis software on a given set of raw data or even navigating the folder tree can in principle be performed manually, this approach does not scale when such operations must be performed repeatedly over long time periods and large datasets.
        This is where an efficient system of shell functionality and tailored aliases can improve productivity while simultaneously reducing the risk of human error.

    \item\label{item:analysis} \textbf{Statistical analysis and reweighting.}
        Raw data can be analysed with the same identical software used in \refcite{Cuteri_2021}, which has now been released in its \texttt{v0.3} as a \texttt{Monte Carlo C++ analysis tools} package~\cite{MCC++}.

        This codebase provides unbiased estimators for moments of arbitrary observables, including \mbox{higher-order} cumulants such as skewness and kurtosis, which are particularly well suited for identifying the nature of the phase transition.
        A key component is the generic implementation of the Ferrenberg--Swendsen multiple histogram method~\cite{Ferrenberg:1989ui}, allowing reweighting in multiple parameters.
        This enables continuous interpolation in the bare parameter space and significantly enhances the efficiency of locating critical points.

        Particular attention has been devoted to the correct implementation of the uncertainty estimate on the reweighted quantities in the multiple histogram method, which is nontrivial in the presence of correlated Monte Carlo data.

        The modularity of the analysis layer has already proven useful beyond the specific workflow discussed in the present work.
        In particular, this analysis package has recently been employed independently for the statistical analysis of lattice QCD data produced within a different simulation framework~\cite{Endrodi:2026tqa}, demonstrating that the analysis infrastructure can be effectively reused outside the original software ecosystem.
        This provides a concrete example that the modularity described above is not merely an architectural design principle: individual components can be detached from the full workflow and reused in independent computational environments.
        This separation between data production and statistical analysis further illustrates the benefits of standardized interfaces and modular software design in large-scale computational physics projects.
\end{enumerate}

The maturity of the framework is supported by its sustained use in large-scale lattice QCD investigations over more than a decade~\cite{Czaban:2015sas,Cuteri:2017gci,Cuteri:2020yke,Philipsen_2020,Cuteri_2021,Kaiser:2021rxe,DAmbrosio:2021awi}, rather than by the proposed application alone.
The same computational principles have therefore been exercised repeatedly in production studies, while the present release brings the corresponding components and interfaces together in a documented and reusable form.
The independent reuse of the Monte Carlo analysis layer described above provides an additional example of this component-level reusability.

A central feature of the presented framework is the full public availability of all software components, combined with a modular design that enforces a clear separation of concerns between simulation, orchestration, and analysis layers, supported by standardized interfaces between these components.
This modularity facilitates independent updates of individual components and enables, in principle, straightforward extensions to different lattice actions, fermion discretizations, or parameter regimes.
Taken together, the described framework constitutes a reproducible computational blueprint for large-scale lattice QCD studies at zero and nonzero imaginary chemical potential, lowering the barrier for future investigations of the QCD phase diagram.
The explicit separation between workflow layers and their interfaces further enables reproducibility across different computing environments and facilitates independent validation of individual components.

\subsection*{Clean software development}

The software components forming the core of the present framework have been developed with a strong emphasis on clean software engineering practices~\cite{martin2009clean}.
In particular, the main codebases, namely \clqcd, \bahamas, and the \texttt{Monte Carlo C++ analysis tools} package, follow a modular design with well-defined interfaces, enabling a clear separation of concerns between simulation, orchestration, and data analysis.
A significant effort has been devoted to ensuring correctness and long-term maintainability through systematic testing.
This includes both unit tests for individual components and integration tests covering the interaction between different modules.
In several parts of the development process, a test-driven development (TDD) approach has been adopted, leading to improved code reliability and facilitating future extensions.
This aspect is particularly relevant in the context of large-scale lattice QCD simulations, where undetected software errors may propagate into costly numerical campaigns.

In contrast, the suite of utility scripts currently released as version \texttt{v0.1} serves a different purpose.
These scripts have been developed over time to support daily research activities, prioritizing flexibility and rapid prototyping over strict architectural design.
As a consequence, they presently exhibit a more coarse-grained structure and partially accumulated technical debt.
However, first steps towards a systematic refactoring have already been undertaken, including the extraction of reusable components and the consolidation of frequently used functionality.
Future development will focus on transforming this collection into a more structured and extensible infrastructure, aligning it with the design principles adopted for the core software components.

This distinction between rigorously engineered core software and more exploratory user-level tooling reflects a pragmatic development strategy.
It enables both high reliability in critical components and sufficient flexibility to adapt to evolving research needs, while maintaining a clear path towards improved software quality and reproducibility.


\section{Conclusions}

\begin{table*}[t]
    \centering
    \begin{tabular}{@{\quad}rl@{\quad}}
        \toprule
        \textbf{\clqcd:} & \url{https://doi.org/10.5281/zenodo.5121917} \\
        \textbf{BaHaMAS:} & \url{https://doi.org/10.5281/zenodo.17980772} \\
        \textbf{Monte Carlo C++ analysis tools:} &  \url{https://doi.org/10.5281/zenodo.17987144}\\
        \textbf{LQCD utility scripts suite:} &  \url{https://github.com/AxelKrypton/LQCD-script-utilities}\\
        \bottomrule
    \end{tabular}
    \caption{%
        The software components discussed in this work are all publicly available and most of them are findable on Zenodo.
    }
    \label{tab:code}
\end{table*}

This work presented a scalable and reproducible computational framework for large-scale lattice QCD investigations at finite temperature and imaginary chemical potential.
The presented software ecosystem brings together GPU-accelerated Monte Carlo simulation codes, automated HPC workflow orchestration, intermediate data-management utilities, and statistically controlled analysis tools into a coherent open-source infrastructure.

From the physics perspective, the framework naturally supports systematic extensions of the investigation introduced in \refcite{Cuteri_2021} towards nonzero purely imaginary chemical potential.
Such studies may help clarifying the structure of the 3D Columbia plot in the continuum limit and the possible existence or disappearance of chiral first-order regions away from $\mu=0$.
In particular, exploratory investigations at intermediate imaginary chemical potential may provide important indications about the fate of chiral critical surfaces approaching the continuum limit, while avoiding the additional numerical complexity associated with the \RW\ plane.

At the computational level, the presented framework represents the consolidation and public release of a software ecosystem that has already been successfully employed over more than a decade in previous large-scale lattice QCD projects.
The present article systematizes this previously operational infrastructure into a coherent and reusable framework, explicitly documenting the interaction between its components and the software-engineering principles underlying its design.
The quantitative scale of the previous production campaigns, summarized in \cref{tab:workflow-scale}, provides practical evidence for the maturity of the workflow beyond the proposed application.
The framework has been exercised over more than a decade in large-scale lattice QCD investigations involving long-running and strongly interconnected Monte Carlo campaigns on heterogeneous HPC systems.
Together with the independent reuse of the analysis layer, these experiences demonstrate that the presented ecosystem is not merely a conceptual architecture, but a computational infrastructure that has already been used and reused in production.

The modular architecture of the framework, based on a clear separation of concerns between simulation, orchestration, data handling, and analysis layers, enables independent evolution of individual components while preserving interoperability through standardized interfaces.
This design significantly improves reproducibility across heterogeneous HPC systems and facilitates the validation and comparison of different simulation codes within a common analysis environment.
The resulting infrastructure is therefore not limited to the specific physical application discussed here, but is intended as a reusable computational blueprint for a broad class of finite-temperature lattice QCD investigations requiring large coordinated simulation campaigns and adaptive parameter refinement.
The practical flexibility of this modular design has already been demonstrated by the independent reuse of the statistical analysis layer in separate lattice QCD studies based on different simulation frameworks~\cite{Endrodi:2026tqa}.

Several natural extensions of the present framework can already be identified.
On the simulation side, additional lattice QCD software packages could be integrated into the workflow infrastructure following the same abstraction principles currently adopted for \clqcd\ and \oqcd.
At the analysis level, the present sequential implementation of the multiple histogram method could be further accelerated through GPU-based parallelization of its computational bottlenecks, potentially yielding substantial performance improvements even on commodity hardware.
More generally, the modular organization of the software ecosystem provides a robust basis for future extensions towards new algorithms, architectures, and lattice discretizations.
In the longer term, these standardized interfaces could also provide a suitable foundation for higher-level automated agents, including AI-assisted systems, capable of coordinating parts of the simulation and analysis cycle.

All software components discussed in this work are publicly available and released as open-source projects (cf.\ \cref{tab:code}), lowering the barrier for future investigations of the QCD phase diagram and supporting transparent and reproducible computational science.


\section*{Acknowledgments}
    I thank \OP\ for the discussion in March 2023 that led to \cref{fig:3D-seculation-2} and I am grateful to him and his collaborators for subsequent conversations that shed light on many different aspects.
    I acknowledge support by the Deutsche Forschungsgemeinschaft (DFG, German Research Foundation) through the \crc\ `Strong-interaction matter under extreme conditions'~--~project number 315477589~--~TRR 211.

\printcredits

\section*{Declaration of usage of generative AI}

During the preparation of this work, the author made use of a generative AI tool (ChatGPT, OpenAI) exclusively for language refinement.
The author has carefully reviewed and edited the output and takes full responsibility for the content of this publication.


\vfill\null 
\bibliography{Literature}


\end{document}